# Photocatalytic $H_2$ generation using dewetted Pt-decorated $TiO_2$ nanotubes – Optimized dewetting and oxide crystallization by a multiple annealing process

by JeongEun Yoo,[1†] Marco Altomare,[1†] Mohamed Mokhtar,[2] Abdelmohsen Alshehri,[2] Shaeel A. Al-Thabaiti,[2] Anca Mazare,[1] and Patrik Schmuki[1,2]*

1 Department of Materials Science, Institute for Surface Science and Corrosion WW4-LKO, Friedrich-Alexander University, Martensstraße 7, D-91058 Erlangen, Germany

2 Chemistry Department, Faculty of Sciences, King Abdulaziz University, 80203 Jeddah, Saudi Arabia Kingdom

† These authors contributed equally

* Corresponding author. E-mail: schmuki@ww.uni-erlangen.de, Tel.: +49-9131-852-7575, Fax: +49-9131-852-7582

## ABSTRACT

In the present work we use $TiO_2$ nanotube arrays, carrying a Pt coating that is optimally dewetted, as a photocatalyst to generate $H_2$. In order to achieve a maximum $H_2$-generation efficiency, on the one hand an ideal thermal dewetting of the Pt layer into nanoparticles is needed that requires an oxygen free heat treatment, and on the other hand an optimal crystallization of the $TiO_2$ nanotubes into anatase with reduced defect density is achieved only by annealing in $O_2$ containing environment. To overcome this issue, we combine adequate reducing and oxidizing conditions in a multiple annealing treatment, and obtain Pt-decorated anatase $TiO_2$ nanotubes showing significantly enhanced photocatalytic $H_2$ generation ability.

## INTRODUCTION

Since the groundbreaking work of Fujishima and Honda,[1] the generation of $H_2$ through photocatalytic water splitting on semiconductors has been widely investigated, as it is perceived as a most promising way for the sustainable generation of energy. A photocatalytic process is in general based on light absorption by a semiconductor that leads to charge carrier generation: electrons are photopromoted to the conduction band, while holes are left in the valence band. Electrons and holes on the respective bands may then migrate to the semiconductor surface and induce red-ox reactions with species in the environment.[2-4]

Among the different photocatalytic materials that have been developed in the last decades, $TiO_2$ still represents one of the most promising semiconductor metal oxides, being cheap, non-toxic, (photo-)chemically stable and, most importantly, having suitable band edge positions. For anatase $TiO_2$, the conduction band edge lies at ca. $- 0.25$ V (vs. SHE, pH 1)[5,6] – the conduction band electrons are thus thermodynamically able to reduce water into $H_2$. Nevertheless, this reaction is not only hampered by charge carrier recombination and trapping phenomena, but also by the sluggish kinetics of electron transfer to the environment.[7] These issues can be tackled by i) nanostructuring the semiconductor,[8,9] and by ii) using suitable co-catalysts (mostly Pt, Pd and Au).[10-14]

When nanostructured $TiO_2$ is used as photocatalyst, the charge carriers have to migrate only over relatively short distances (nm-scale) to reach the surface and react with the environment, and this reduces the probability of charge carrier recombination and trapping.[15] Among several approaches to fabricate nanostructures,[8,16-18] electrochemical anodization of metallic Ti, that leads under self-organizing conditions to highly-ordered and vertically aligned $TiO_2$ nanotubes (NTs),[19,20] is one of the most facile techniques. A key advantage of this approach is that the structural properties of the nanotubes (wall thickness, tube diameter and length, etc.) can be easily tailored simply by the electrochemical conditions (e.g., adjusting applied voltage, water and fluoride contents).[21,22]

The second typical measure to increase the efficiency of $TiO_2$ photocatalysts is the deposition of noble metal co-catalysts, such as Au, Pd or Pt, onto its surface. These co-catalysts not only act as "electron sink" (i.e., by trapping conduction band electrons),[23,24] but also enable efficient electron transfer to the environment.[6]

Recently we reported on a most straightforward way to decorate $TiO_2$ nanotube layers with co-catalyst particles that is to first coat the tubes with a thin sputter-deposited film of the noble metal (Au, Pt) and then to thermally dewet this thin film to particles,[25-28] i.e., a suitable thermal treatment breaks up the metal film into small particles.[29] Dewetting of thin Au films on $TiO_2$ can be achieved by a simple annealing in air at 450°C. These annealing conditions are ideal as they also lead to crystallization of the $TiO_2$ tubular substrate (that after formation is amorphous) into anatase which is the most photocatalytically active $TiO_2$ polymorph. In other words, a single step annealing can at the same time be used to crystallize $TiO_2$ to anatase and induce Au dewetting.

While this air treatment is ideal for Au, for Pt (an even more effective co-catalyst)[23] the situation is different. A main issue of Pt is that dewetting on $TiO_2$ substrates normally requires relatively higher temperatures (T ≥ 500°C) and must be carried out in Ar or $N_2$ atmosphere (one of the main reasons is that the occurrence of Pt oxide[30,31] interferes with dewetting and leads to significant photoactivity decay).[32] On the other hand annealing in such a reductive atmosphere leads to a partial reduction of $TiO_2$ – these reduced states lead in general to charge carrier trapping and recombination.[33]

In this work we try to overcome this issue by combining reducing and oxidizing annealing treatments. We show that optimized annealing can lead to a drastically increased photocatalytic efficiency in terms of a maximized $H_2$ evolution efficiency.

## EXPERIMENTAL

Ti foils (0.125 mm thick, 99.7% purity, Goodfellow, England) were degreased by sonicating in acetone, ethanol, deionized water, and then dried in a $N_2$ stream. The Ti foils were then anodized to fabricate the highly ordered $TiO_2$ nanotubes arrays in an electrolyte based on o-$H_3PO_4$ with 3 M HF (Sigma-Aldrich).[25] For the anodic growth, a two-electrode configuration was used, where the Ti foil and a Pt gauze were the working and counter electrodes, respectively. The anodization experiments were carried out applying a potential of 15 V (for 2 h) provided by a Volcraft VLP 24 Pro DC power supply. After the anodization, the nanotube films were rinsed with ethanol and dried in a $N_2$ stream.

The tube oxide layers, either as-formed (i.e., amorphous) or thermally treated, were coated by thin Pt films with a nominal thickness in the range of 1-15 nm.

The tubes were exposed to different thermal treatments. Annealing in ambient air was carried out at different temperatures (350-550°C) for 1 h, using a Rapid Thermal Annealer (Jipelec Jetfirst 100 RTA), with a heating and cooling rate of 30°C min-1. Annealing in pure $O_2$ (99.95 %, Linde) was carried out at 450°C for 30 min, in a VMK80S Linn High Therm tubular furnace. The $O_2$ flux was set to 100 mL min-1. Annealing in pure $N_2$ (99.999 %, Linde) was carried out at 600°C for 1 h, in a ZEW 1041-5 Heraeus tubular furnace. The $N_2$ flux was set to 20 NL min$^{-1}$.

The deposition of Pt on the anodic $TiO_2$ films was carried out using a high vacuum sputter coater (Leica – EM SCD500). The pressure of the sputtering chamber was reduced to $10^{-4}$ mbar, and then set at $10^{-2}$ mbar of Ar. The applied current was 15 mA. The amount of sputtered Pt was determined by an automated quartz crystal film-thickness monitor.

For morphological characterization, a field-emission scanning electron microscope (Hitachi FE-SEM S4800) was used. X-ray diffraction analysis (XRD, X'pert Philips MPD with a Panalytical X'celerator detector) using graphite monochromized Cu Kα radiation (wavelength 1.54056 Å) was used for determining the crystallographic composition of the samples.

The composition and the chemical state of the films were characterized using X-ray photoelectron spectroscopy (XPS, PHI 5600, US), and the spectra were shifted in relation to the C1s signal at 284.8 eV (Pt4f peaks were fitted with Multipak software).

The photocatalytic $H_2$ generation experiments were carried out by irradiating the oxide films with UV light (HeCd laser, Kimmon, Japan; $\lambda$ = 325 nm, expanded beam size = 0.785 cm$^2$, nominal power of 60 mW cm$^{-2}$) in a 20 vol% ethanol–water solution (for 9 h) in a quartz tube. The amount of produced $H_2$ (which accumulated over time within the tube) was measured at the end of the experiments by using a gas chromatograph (GCMSQO2010SE, Shimadzu) equipped with a thermal conductivity detector and a Restek micropacked Shin Carbon ST column (2 m × 0.53 mm). GC measurements were carried out at a temperature of the oven of 45°C (isothermal conditions), with the temperature of the injector set at 280°C and that of the TCD fixed at 260°C. The flow rate of the carrier gas, i.e., argon, was 14.3 mL min$^{-1}$. Before the photocatalytic experiments, the reactor and the water-ethanol mixtures were purged with $N_2$ for 30 min to remove $O_2$. This is strictly needed as $O_2$ would diminish the efficiency of $H_2$ generation by competitively undergoing photocatalytic reduction to $O^{2-\bullet}$ (i.e., $O_2$, instead of water or ethanol, is reduced by conduction band electrons).[3,34] The photocatalytic experiments were carried out in ethanol–water solution since the presence of specific amounts of organics significantly triggers the $H_2$ generation. Ethanol acts in fact as a hole-scavenger, that is, the organic molecules are quickly oxidized towards several intermediates and eventually to $CO_2$. As a result of the fast hole-consumption, conduction band electrons are more readily available for the $H_2$ generation reaction.[35]

**RESUTLS AND DISCUSSION**

Fig. 1 shows the morphological features of the short-aspect anodic $TiO_2$ nanotube layers used in this work. The nanotubes are almost ideally hexagonally packed and have average diameter and length of 70-80 nm and 150-170 nm, respectively.[25] Fig. 1 (b,c) show morphology of the tube substrates after sputter-coating the tubes with a layer of Pt with a nominal thickness of 5 nm. The profile of the noble metal deposition is particularly clear in Fig. 1 (c), where one can see that the Pt film preferentially coats the tube top, and the thickness decreases gradually toward the bottom of the cavities.

The thin Pt films present on the tube substrates can be split up into arrays of fine particles by a thermal treatment at 600°C in $N_2$ atmosphere for 1 h,[36] leading to the Pt/$TiO_2$ structures shown in Fig. 1 (d,e). Here one can see that in the case of 5 nm-thick sputter-coated Pt films, the particles formed by dewetting are globular and have an average diameter of ca. 5-25 nm, with smaller particles at the bottom and bigger aggregates at the top of the tubes. Such a distribution of the nanoparticles (NPs) is in line with the different initial thickness of the Pt film on the tube walls, that is, the size of the dewetted particles strictly depends on the thickness of the metal film and thinner films generally split into small particles, while thicker coatings form large metal islands.[29]

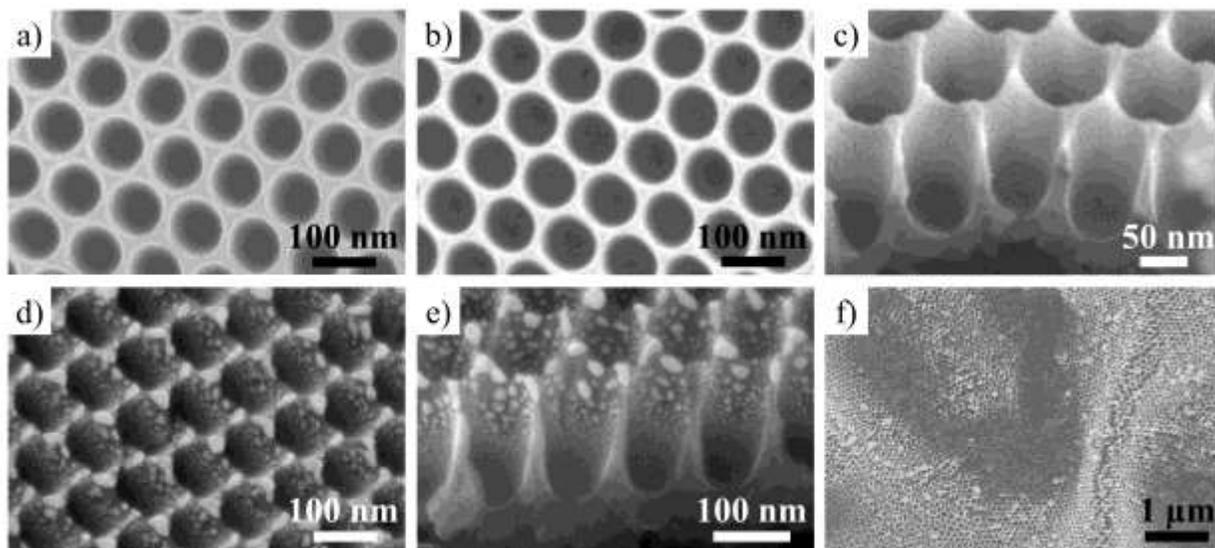

Figure 1. SEM images of TiO$_2$ nanotube arrays a) as-formed; b) and c) sputter-coated with a 5nm-thick Pt film; d) and e) after thermal dewetting of the Pt film by heat treatment in N$_2$ at 600ºC for 1 h; f) exposed to a thermal treatment in air at 600ºC for 1 h.

Noteworthy, only a thermal treatment at 600°C (or higher) in N$_2$ (or Ar) could lead to defined Pt dewetting onto TiO$_2$ surfaces while lower temperatures did not affect the morphology of the Pt films even after extended times. A treatment in an O$_2$-containing atmosphere at 600°C not only leads to partial dewetting of Pt (presumably due to Pt oxide formation) but, even more detrimental, leads also to a complete collapse of the tube structure as shown in Fig. 1 (f). An oxygen treatment at 600°C also leads to extensive growth of thermal titanium oxide from the Ti metal substrate underneath the tubes.[24,28,37-39]

A first round of photocatalytic experiments showed that Pt dewetting is highly beneficial for the photocatalytic H$_2$ generation even when this process is carried out in N$_2$ at 600°C. Dewetted films showed a H$_2$ production of ~ 20 µL of H$_2$ / 9h in comparison to non-dewetted layers that produced ~ 10 µL / 9h, that is, a 100% enhancement can be achieved through the dewetting step.[25,27]

This result is on the one hand most likely ascribed to the fact that as-sputtered noble metal films coat the $TiO_2$ tubes in a conformal fashion, and thus shade the underneath structure limiting the absorption of UV light by the semiconductor. On the other side, dewetting forms arrays of Pt nanoparticles homogeneously distributed on the oxide surface and with higher specific area (than a coherent Pt film) that improves the electron transfer to the environment and, as a consequence, the $H_2$ generation efficiency.[26]

Nevertheless, Pt-coated tubes annealed in air at 450°C show an even higher $H_2$ production of ~ 60 µL / 9h, although this thermal treatment does not induce Pt dewetting. This means that an enhanced $H_2$ evolution must rather be related to a higher quality oxide in the tubes (lower defect density anatase formed by crystallizing in non-reductive conditions).[40]

This clearly indicates that both Pt dewetting (high temperature treatment in $N_2$) and a suitable crystallization of the oxide (air annealing) are highly beneficial to achieve an enhanced photocatalytic activity. In order to optimize both benefits we explored several annealing conditions of the Pt-coated tubes, at various temperatures and in different atmospheres, and investigated the properties of the films in view of improving their photocatalytic $H_2$ generation ability.

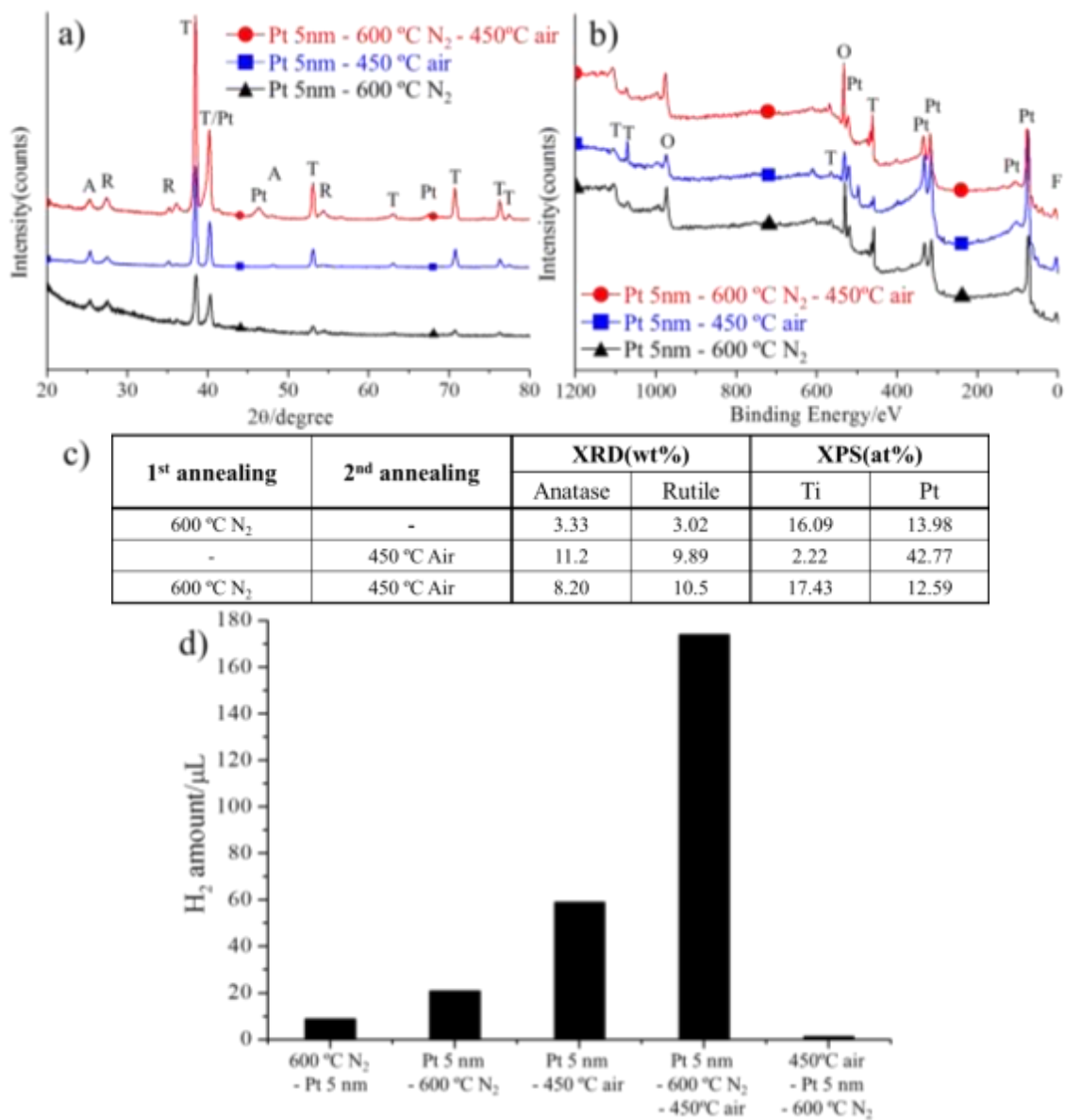

Figure 2. a) XRD patterns, b) XPS surveys c) summary of the XRD and XPS results, and d) photocatalytic $H_2$ generation results of Pt/TiO$_2$ nanotube layers exposed to different thermal treatments.

Fig. 2 (a-c) show the XRD and XPS results of differently treated samples. In every case we firstly deposited on the tube layers 5 nm-thick Pt films, and then exposed the layers to different thermal treatments. From XRD it is evident that for tubular oxides treated in $N_2$ at 600°C only a relatively low degree of crystallinity is obtained. The content of anatase and rutile $TiO_2$ phase is only 3.3 and 3.0 wt%, respectively. The low degree of crystallinity can be ascribed to the amorphous nature of as-formed anodic oxides containing a large density of oxygen vacancies – a defect free crystalline material requires annealing in an $O_2$-containing atmosphere.[33,41]

In fact by air-annealing at 450°C, the content of anatase and rutile in the oxide films increased up to ca. 11.2 and 9.9 wt%, respectively.[42] However, as mentioned before, air-annealing does not induce Pt dewetting. Nevertheless, we found that a multiple treatment consisting of $N_2$-annealing at 600°C followed by air-annealing at 450°C can provide dewetting and leads as well to a relatively high degree of oxide crystallinity, with contents of anatase of ca. 8.2 wt%.

XPS results showed all films to be generally composed of $TiO_2$ and Pt metal, with a small content of fluorine, most probably due to F-ion uptake by the oxide during the electrochemical growth in the HF electrolyte. Also, from the XPS data the occurrence of Pt dewetting is traceable. The surface Ti amount apparent in XPS for non-dewetted films (air-annealed at 450°C) is relatively low (2.2 at%) while Pt is accordingly high (42.8 at%), confirming that the oxide surface is well coated by the sputtered Pt film. After the films are $N_2$-treated at 600°C the Ti surface content increased up to 16-17 at%, indicating that Pt agglomerated and thus more open oxide surface becomes detectable.

The Pt/oxide films treated under different annealing conditions were tested as catalysts for $H_2$ generation from $H_2O$-ethanol solutions. From the results summarized in Fig. 2 (d) it is clear that the different thermal treatments (different temperatures and atmospheres) have dramatic effects on the $H_2$ generation efficiency.

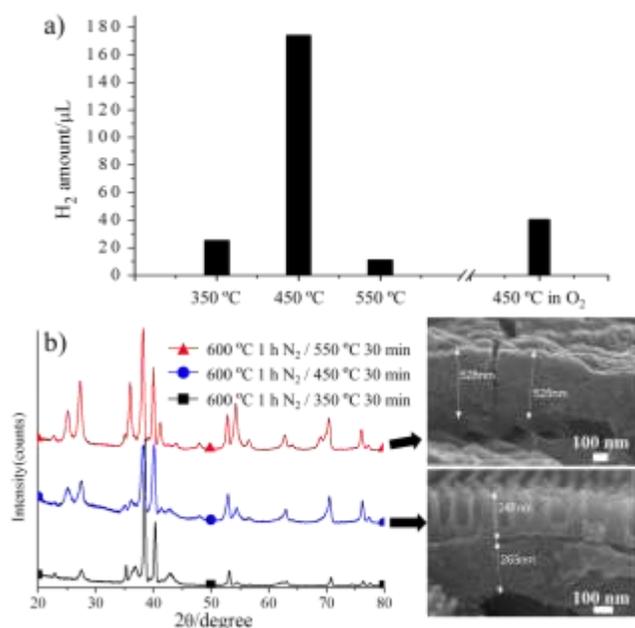

Figure 3. a) Photocatalytic $H_2$ generation data and b) XRD patterns of $TiO_2$ nanotube layers decorated with sputter-dewetted particles and exposed to different thermal treatment in $O_2$-containing environment. The SEM images in b) show the thickness of the rutile under layer formed after air-annealing at 450ºC (lower) and 550ºC (upper).

In line with the observations above, a first level of photocatalytic enhancement is achieved with the more co-catalytically efficient Pt nanoparticles formed by dewetting ($N_2$-annealing), compared to a conformal noble metal film (i.e., ~ 20 vs. 10 µL $H_2$ / 9h). Then, the $H_2$ generation efficiency is further improved up to ~ 60 µL $H_2$ / 9h, even without Pt dewetting, by optimizing the quality of the oxide with an air-annealing treatment.[33,43] However, most relevant is that the two concepts can be beneficially combined through a multiple annealing approach, i.e., to provide both Pt dewetting and anatase nanotubes: Pt/tube layers firstly $N_2$-dewetted (600°C) and then again air-treated at 450°C lead to ~ 180 µL $H_2$ / 9h, i.e., the highest $H_2$ generation efficiency among the differently treated Pt/$TiO_2$ photocatalysts.

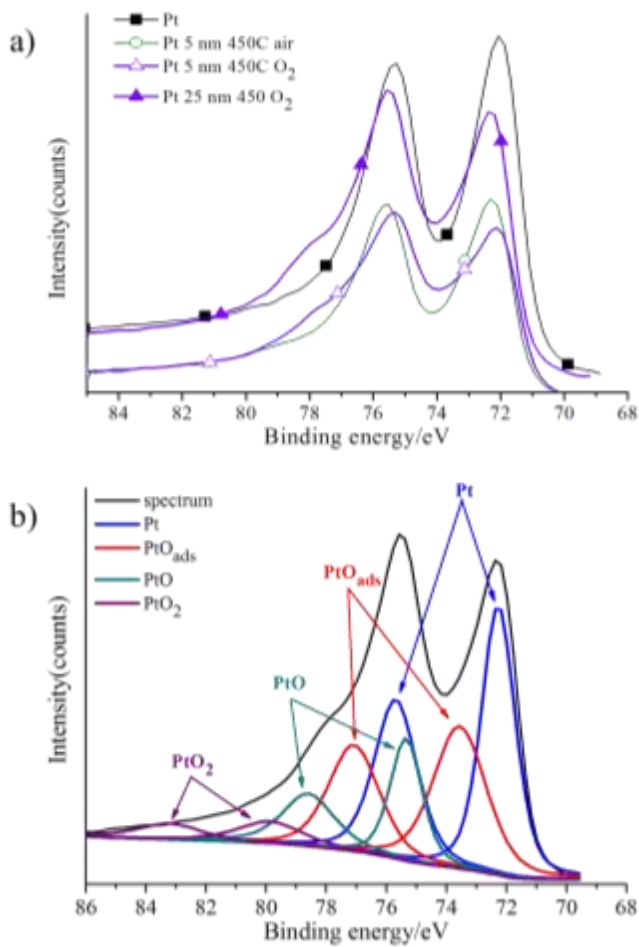

Figure 4. a) High-resolution Pt4f XPS spectrum of Pt (reference) and $TiO_2$ nanotube layers decorated by Pt sputter-dewetting and exposed to different thermal treatments in $O_2$-containing atmosphere. The Pt4f spectrum of the layer sputter-coated with a 25 nm-thick Pt film is shown in b) along with the fitting and deconvolution of contributions from PtOads, PtO and $PtO_2$.

Interestingly, if the sequence is inverted, i.e., Pt is dewetted ($N_2$/600°C) only after crystallizing the tubes in air, the $H_2$ evolution drops to ~ 1 μL / 9h. This is ascribed to generation of oxygen vacancies during the second annealing in reductive conditions. Such reductive defect generation is widely reported in the literature.[36,37,44-47]

The air-crystallization step was further explored by exposing dewetted layers to air-annealing at different temperatures in the 350-550°C range. The photocatalytic results summarized in Fig. 3 (a) show that compared to an air-treatment at 450°C (~ 180 µL $H_2$ / 9h), annealing at 350 and 550°C leads to rather low $H_2$ generation efficiencies (~ 20 and ~ 10 µL $H_2$ / 9h, respectively). The reasons for this become evident from the XRD data (Fig. 3 (b)). One can see that the air/350°C treatment does not induce oxide crystallization into anatase $TiO_2$ – only a small rutile peak is visible that may even be ascribed to the rutile formed with the previous annealing step (dewetting in $N_2$). Air-annealing at 550°C on the other hand leads to higher degree of crystallinity and to anatase formation, but at the same time forms a large amount of rutile.[22,48]

This rutile formation at 550°C can be ascribed to thermal oxidation of the Ti metal substrate. In line with other works, the growth of rutile occurs firstly by rutile seeding at the metal/$TiO_2$ interface and then proceeds (with higher annealing temperatures or longer thermal treatments) towards the tubes top.[24,37,42,49] This is visible from the cross-sectional SEM images in Fig. 3 (b). Here one can see that rutile forms as a layer of some hundreds of nanometers between the tubular oxide and the Ti metal substrate. An air-treatment at 550°C leads however to a ca. 500 nm-thick rutile film, that is 2-time the thickness of that formed at 450°C.

Therefore, the absence of anatase phase in the layers air-treated at 350°C, and the predominant rutile content in the oxides crystallized at 550°C, are the reasons for the small $H_2$ generation yield.

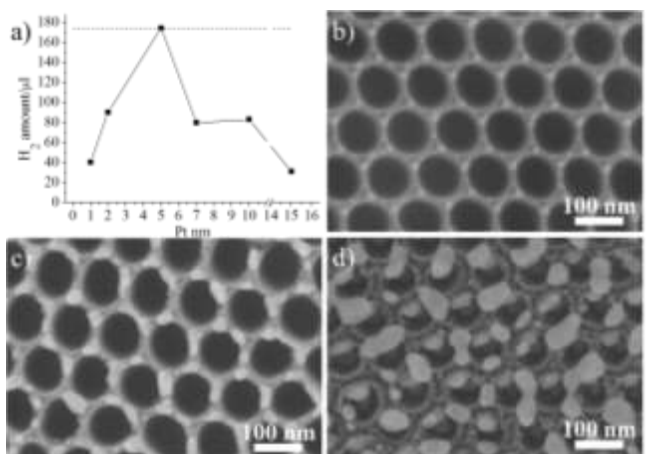

Figure 5. a) Photocatalytic $H_2$ generation results, and b), c), and d) SEM images of $TiO_2$ nanotube layers decorated by sputter-dewetting different Pt amounts, such as b) 2 nm, c) 7 nm, and d) 10 nm.

We also found that annealing in $O_2$-containing atmosphere can, if not properly adjusted, affect the Pt oxidation state[30] and therefore the co-catalyst efficiency.[32] Pt/$TiO_2$ nanotubes annealed after dewetting in pure $O_2$ at 450°C produced ~ 40 µL $H_2$ / 9h. Please note that films treated at the same temperature in air lead intstead to ~ 180 µL $H_2$ / 9h.

From the XPS data in Fig. 4 it can be seen that the noble metal in films air-treated at 450°C consists of metallic Pt (i.e., $Pt^0$),[30] and the fitting of the HR spectrum in the Pt4f region shows besides the $Pt^0$ signals only a small contribution (small shoulder) of adsorbed oxygen. On the contrary, for films annealed in pure $O_2$ an inversion of the relative intensity of the Pt4f signals is evident, and even more importantly a broad shoulder (at ~ 76-80 eV) appears that can be attributed to the formation of PtO (PtII) and $PtO_2$ (PtIV).[30]

To explore more in details the possibility of Pt oxidation by $O_2$-annealing, Pt/tube films sputter-coated with significantly thicker Pt films (25 nm) were exposed to an identical multiple annealing ($N_2$/600°C followed by air/450°C). The HR XPS spectrum in Fig. 4 (a) shows an even more pronounced shoulder, and its fitting (Fig. 4 (b)) matches well with the reference Pt4f signals of PtII and PtIV oxides.[50,51] The low $H_2$ generation efficiency of tube layers annealed in pure $O_2$ is thus ascribed to the formation of Pt oxides which deteriorates its efficiency as a co-catalyst.[31,32, 50, 51]

In addition to the thermal treatment, we also evaluated the effect of the amount of co-catalyst on the $H_2$ evolution efficiency, after optimized Pt dewetting and oxide crystallization. The photocatalytic results in Fig. 5 (a) show a clear enhancement of the $H_2$ generation when the amount of co-catalyst is increasing up to 5 nm. Larger co-catalyst amounts, e.g., 7-15 nm-thick Pt films, showed instead a much lower photocatalytic activity.[26]

From the SEM images in Fig. 5 (b-d) it is evident that an increase of the Pt film thickness leads to an increase of the average Pt nanoparticle size after dewetting. For Pt film of 2, 7 and 10 nm, the average Pt NPs size ranges around 5-20, 30-70 and 80-100 nm, respectively.

The smaller nanoparticles (e.g., Fig. 5 (b)) are round in shape, show a relatively narrow size distribution, and are ordered at the tubes top in hexagonal networks. Conversely, Pt islands as large as several tens of nanometers (formed from thick metal films, e.g., Fig. 5 (d)) show undefined shape and a rather broad size distribution.[29]

The reason for the drop of photocatalytic activity observed when increasing the co-catalyst amount is therefore that while dewetting of thin Pt films exposes well the underneath oxide to the environment, thick noble metal films dewet into large Pt islands at the tube top that shade the underneath oxide and limit light absorption, this resulting in a reduced density of photo-promoted electrons. Additionally, thin metal films form Pt nanoparticles with higher specific surface area than the large Pt islands and this enhances the area for electron transfer to the environment and therefore the $H_2$ evolution efficiency.

**CONCLUSION**

In this work we demonstrate for $TiO_2$ nanotubes that the quality of the oxide, in terms of crystallinity and defect density, and the morphology and chemical state of the co-catalyst (here Pt) play a crucial role in determining the photocatalytic efficiency of noble metal/$TiO_2$ systems. We show that a sputter-dewetting approach is a suitable technique for decorating anodic nanotube $TiO_2$ surfaces with fine co-catalyst nanoparticles. Even more importantly, we introduced a multiple annealing strategy by combining adequate reducing and oxidizing conditions that led to Pt dewetting and at the same time to the oxide crystallization into anatase nanotubes with reduced defect density. As a consequence, we achieved a maximized $H_2$ generation from such Pt/$TiO_2$ photocatalysts. This concept can be likely extended to other functional metal/oxide systems where the metal is even more susceptible to the annealing conditions.

**ACKNOWLEDGMENT**

The authors would like to acknowledge the ERC, the DFG, and the DFG "Engineering of Advanced Materials" cluster of excellence for financial support. This project was funded by the Deanship of Scientific Research (DSR), King Abdulaziz University, under grant no. 16-130-36-HiCi.

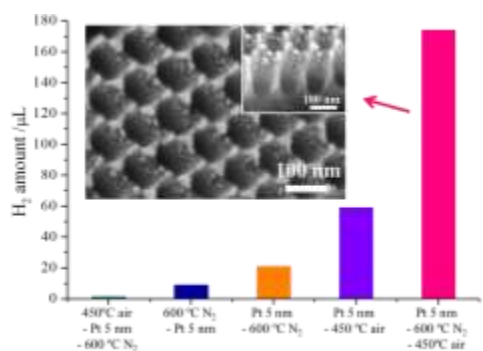